%% file: pic_proceedings.tex
\pdfoutput=1

\documentclass[twocolumn,twoside,slac_two]{revtex4}
\usepackage{graphicx}
\usepackage{fancyhdr}
\usepackage{graphics}
\usepackage{epstopdf}
\usepackage{textpos}
\input{mycmds}

\begin{document}

\title{\centering Muon Neutrino Disappearance and Tau Neutrino Appearance}


\author{
\centering
\begin{center}
M.C. Sanchez
\end{center}}
\affiliation{\centering Iowa State University, Iowa, 50010, USA}
\begin{abstract}
Since evidence for neutrino oscillations was first observed in 1998, the study of muon neutrino oscillations has been aggressively pursued. In doing so, atmospheric and accelerator-based neutrino experiments have measured with the highest precision two fundamental neutrino parameters: the mass-square difference and the large mixing angle in the atmospheric neutrino sector. Furthermore, the dominant mode of these oscillations has recently been established to be from muon to tau neutrinos with both direct and indirect observations. Also, for the first time the anti-neutrino counterparts to these oscillation parameters are being studied. While a consistent picture of the mu-tau sector is thus emerging, a new generation of accelerator-based experiments using off-axis neutrino beams to access this sector could lead to new discoveries.
\end{abstract}

\maketitle
\thispagestyle{fancy}

\section{INTRODUCTION}

The phenomenon of neutrino oscillations, in which neutrinos change flavors as a consequence of having non-zero mass, has been solidly established over the last decade by detecting neutrinos from the atmosphere and from the sun~\cite{ref:atmosnu, ref:solarexp} and by using artificial sources of neutrinos such as accelerators and reactors~\cite{ref:accel,ref:react}. These observations support the description of neutrinos in two distinct flavor and mass bases: 

\begin{equation}
\ket{\nu_\alpha} = \sum_i U_{\alpha i} \ket{\nu_i},
\end{equation}
where the $\alpha = e, \mu, \tau$ labels the electro-weak flavor states and $i =
1, 2, 3$ labels the mass states.

The neutrino states are related by the PMNS neutrino mixing matrix~\cite{ref:PMNS}, where the measured mixing angles have been found to be large, in contrast to the quark sector.  This  matrix can be parametrized as follows: 
\begin{eqnarray}
U 
 & = &
\left(
 \begin{array}{ccc}
  1 &  0      & 0      \\
  0 &  c_{23} & s_{23} \\
  0 & -s_{23} & c_{23} \\
 \end{array}
\right) \times
\left(
 \begin{array}{ccc}
  c_{13}                 & 0 & s_{13} e^{-i \delta_{CP}} \\
  0                      & 1 & 0                     \\
  -s_{13} e^{i \delta_{CP}} & 0 & c_{13}                \\
 \end{array}
\right) \nonumber \\  &\times &
 \left(
 \begin{array}{ccc}
  1 &  0      & 0      \\
  0 &  c_{12} & s_{12} \\
  0 & -s_{12} & c_{12} \\
 \end{array} 
\right), 
\label{eqn:mns}
\end{eqnarray}
where $s_{ij}=\sin \theta_{ij}$ and $c_{ij}=\cos \theta_{ij}$,
$\theta_{ij}$ is the mixing angle of $\nu_i$ and $\nu_j$ and $\delta_{CP}$ is a CP-violating phase.

The transitions from one flavor of neutrino to another can be accurately described by a probability that depends not only on the distance traveled by the neutrino ($L$) and its energy ($E$) but also by the difference in squared masses, $\Delta m^2_{ij} \equiv m^2_i - m^2_j$ and the mixing angle $\theta_{ij}$ for the mass eigenstates $\nu_1$, $\nu_2$ and $\nu_3$.  The data naturally splits into two regimes due to the difference of two orders of magnitude in the scale of  the two known \dmsq{}. The first regime applies to solar \nue\ and reactor  \anue\, and it  is driven by the (\dmsq{21}, \tonetwo) mixing parameters. The second regime describing atmospheric and accelerator results as \numu\ and \anumu\ disappearance is driven by  (\dmsq{32}, \ttwothree). For the results presented here, we are concerned only with this second regime, which can be studied via the 2-flavor approximation of the survival probability: 

\begin{equation}
P (\numu \rightarrow \numu) \approx
1 - \sin^2 2\theta_{23} \sin^2(1.27 \Delta m^2_{32} \frac{L}{E}).
\end{equation}

Neutrino experiments in recent years, notably results from MINOS~\cite{ref:minos-cc-prl} and Super-Kamiokande~\cite{ref:superk} have provided unprecedented precision for two of the fundamental neutrino mixing parameters:  (\dmsq{32} and \ttwothree) (corresponding to the atmospheric sector) by observing muon neutrino disappearance, $i.e.$ the probability \numunumu. These experiments have also addressed the differences that might exist in these parameters when considering anti-neutrinos. Furthermore, the dominant mode in this regime has been shown to be \numunutau\ by OPERA~\cite{ref:opera-nim} and Super-Kamiokande. 

In these proceedings, we describe recent results in muon neutrino disappearance by the MINOS and Super-Kamiokande experiments which measure \dmsq{32} and \ttwothree\ with the highest precision. Next we discuss the tau neutrino appearance results by OPERA which observes this phenomenon directly and again Super-Kamiokande which in this case provides a statistical measurement. In a similar manner we show that MINOS provides a direct measurement of muon anti-neutrino disappearance while Super-Kamiokande makes an indirect measurement using constraints in the atmospheric flux and the neutrino measured parameters. Finally, the new generation of long-baseline experiments, T2K~\cite{ref:t2k-nim} and NOvA~\cite{ref:nova-tdr}, also seek to measure \numu\ disappearance in the very narrow spectrum of their off-axis beams. We present recent results from T2K and future sensitivities for NOvA. For each experiment, we offer a brief description the first time it is mentioned.

\section{MUON NEUTRINO DISAPPEARANCE}

\subsection{The MINOS Experiment} \label{sec:MINOS}

\begin{figure}[t]
\centering
\includegraphics[width=75mm]{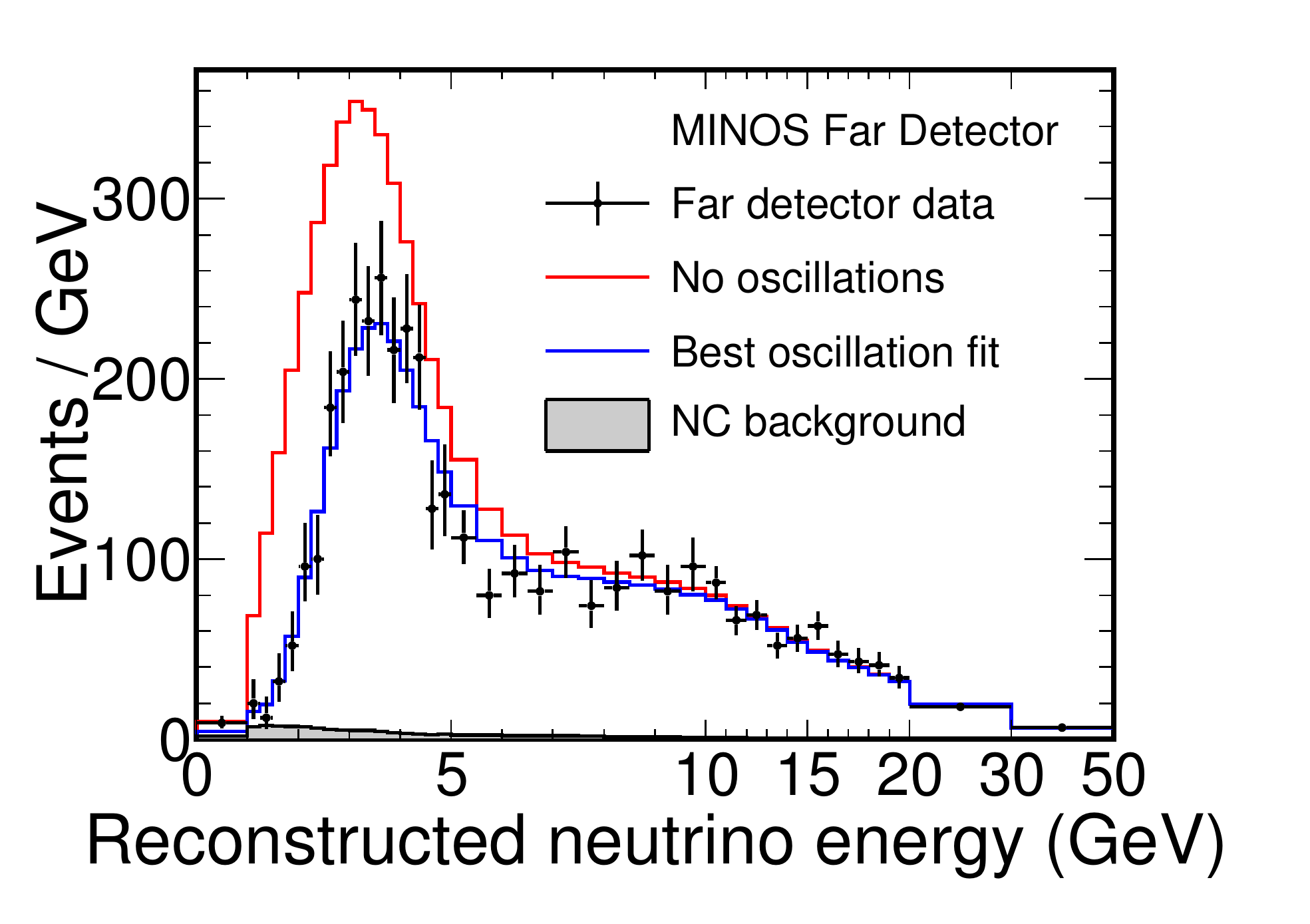}
\includegraphics[width=75mm]{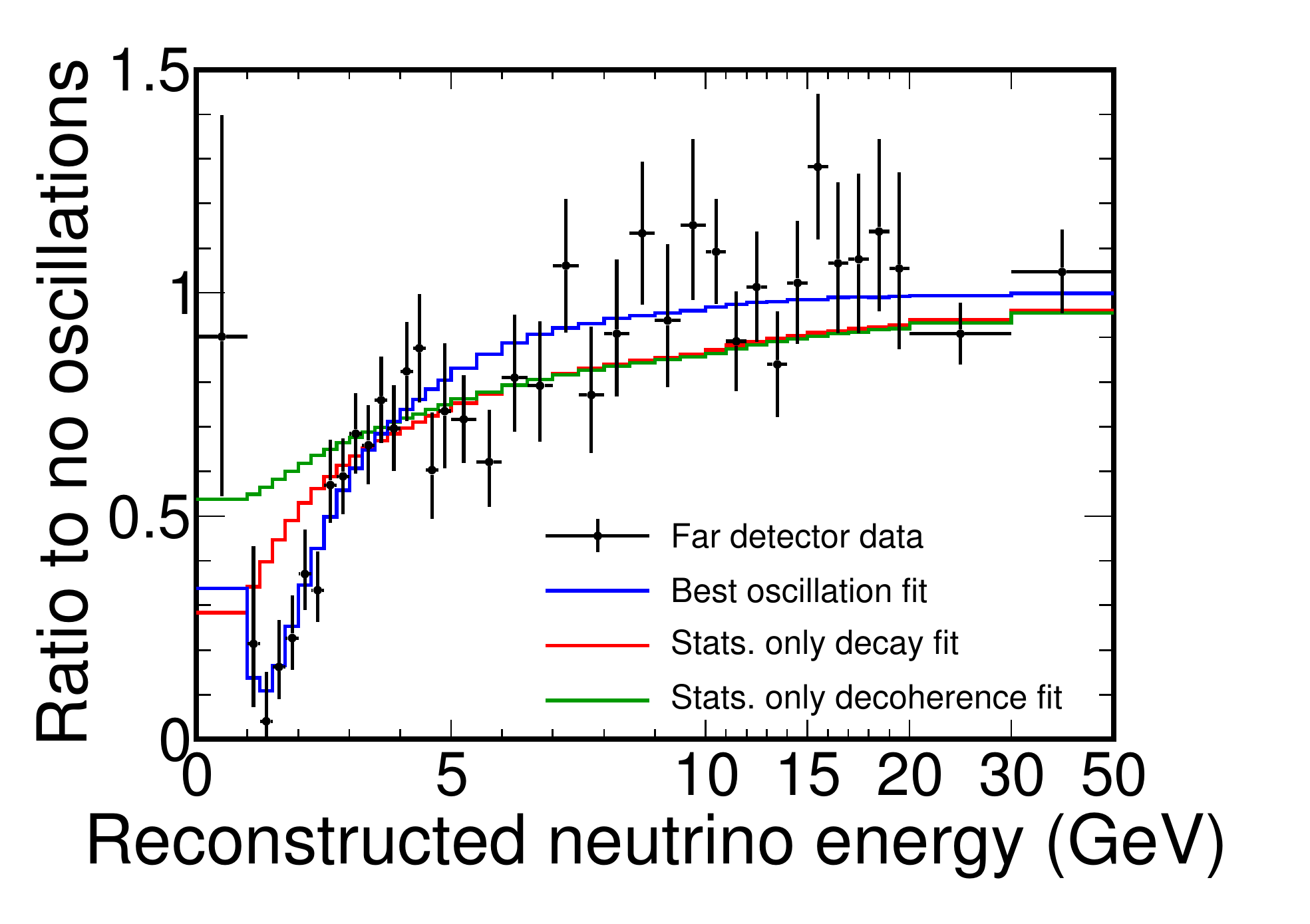}
\caption{Top: The energy spectra of  fully reconstructed events in the MINOS far detector classified as charged current interactions. It shows the MINOS data compared with the non-oscillation spectrum predicted from near detector data (light), the best fit of the oscillation hypothesis (dark) and the predicted neutral current background (shaded).
Bottom: The points with error bars are the background-subtracted ratios of data to the no-oscillation hypothesis.  Lines show the best fits for: oscillations, decoherence and decay.
} \label{fig:minos-spectra}
\end{figure}

MINOS is a long-baseline neutrino oscillation experiment tasked to make a precision measurement of the neutrino mixing parameters associated with the atmospheric neutrino mass splitting. Using a high powered neutrino beam from the Main Injector (NuMI) facility at Fermilab, it compares the neutrino energy spectrum for neutrino interactions observed in two large detectors located at Fermilab and in the Soudan mine in northern Minnesota.

The neutrino beam is produced by impacting 120~GeV protons from the Fermilab Main Injector upon a graphite target, producing pions and kaons, which are then focused by two magnetic horns~\cite{ref:numi}.  The horn current and position of the target relative to the horns can be configured to produce different neutrino energy spectra. Most of the physics data has been acquired in the low energy configuration with a peak at 2-3~GeV, but other configurations have been used to study the backgrounds and systematics. The beam in neutrino mode is mostly \numu{} with small 7\%  \anumu{} and 1.3\% \nue+\anue components. The beam can also run in a mode with an enhanced anti-neutrino component. 

In MINOS, neutrino interactions from the Fermilab Main Injector (NuMI) facility~\cite{ref:numi} are recorded at two detectors: Near Detector (ND), 1~km from the NuMI target, and a Far Detector (FD) at 735~km from the same target. The detectors are magnetized tracking calorimeters composed of 2.54~cm thick steel absorber planes and 1.0~cm thick active scintillator planes composed of 4.1~cm wide strips.  The data recorded in the ND establishes the properties of the beam before oscillations have occurred~\cite{ref:minos-cc-prl,ref:minosnim, ref:minosprd}. Evidence for oscillations is observed as distortions of the beam spectrum or composition measured at the FD with respect to the ND. Thanks to the similarity between the two detectors, systematic errors arising from uncertainties in the neutrino interactions or in the neutrino flux largely cancel. 

MINOS has recorded an exposure of $7.25 \times 10^{20}$ protons-on-target (POT) in neutrino mode. The data selected for the \numu\ disappearance analysis consists of a pure sample of \numu\ charged current (CC) events. The selection is done using a k-nearest-neighbor technique based on variables related to the muon track properties~\cite{ref:minosCC2010}. The neutrino energy is reconstructed by summing the muon track momentum (determined from range or curvature measurements) and the hadronic shower energy. 

We observe a total number of 1986 fully-contained events in the far detector, whereas the expectation without oscillations is 2451 such events. Figure~\ref{fig:minos-spectra} shows the energy spectrum of the selected contained events compared to the non-oscillation prediction and the best oscillation fit which uses the two-neutrino flavor approximation for the survival probability shown earlier. This figure also shows the ratio of the observed energy spectrum to the prediction for null oscillations, the best oscillation fit and fits to alternative models: pure decoherence~\cite{ref:decoherence} and pure decay~\cite{ref:neutrino-decay}.  Additionally we select a non-contained sample (not shown) where we expect 2206 events and observe 2017 as predicted by oscillations. By including this sample we disfavor pure decoherence relative to oscillations at more than 9$\sigma$ and similarly, we disfavor pure decay at 7$\sigma$. 

\begin{figure}[t]
\centering
\includegraphics[width=80mm]{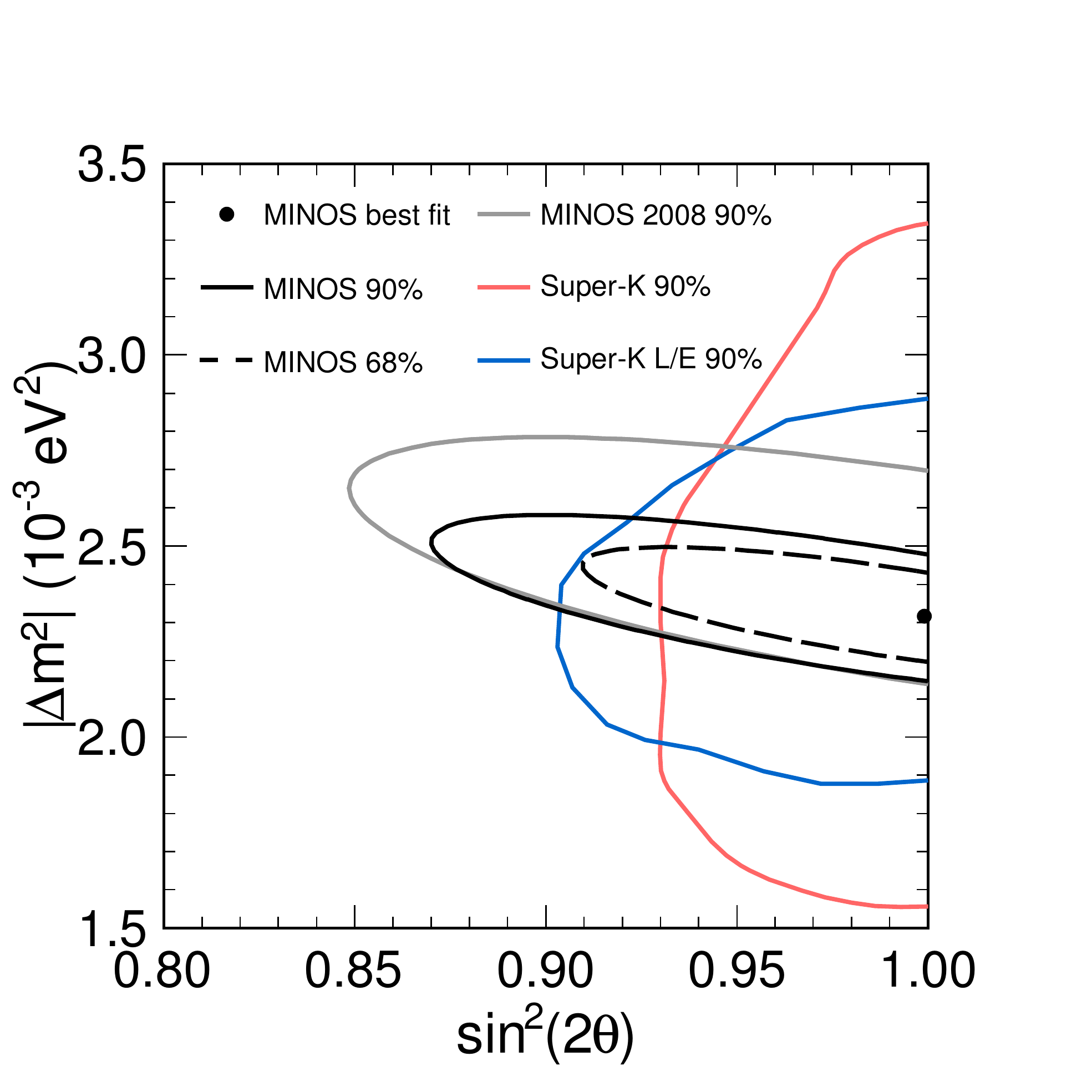}
\caption{MINOS likelihood contours of 68\% (dashed) and 90\% (solid) C.L. around the best fit values  for the mass splitting and mixing angle. Also shown are contours from previous measurements~\cite{ref:minos-cc-prl,ref:superk}.} \label{fig:minos-contours}
\end{figure}

MINOS measures with the World's highest precision the mass-squared difference at  $|\Delta m_{32}^2| = (2.32^{+0.12}_{-0.08})\times10^{-3}$\,eV$^2$. For the mixing angle the best fit is $\rm sin^2\!(2\theta_{23})=1.00$ (when constrained to be physical). The limits for this parameter are  $\rm sin^2\!(2\theta_{23})  > 0.94 ~(0.90)$  at 68~(90)\% confidence level (C.L.). The best fit values with the resulting 68\% and 90\%\,C.L. contours are shown in Fig.~\ref{fig:minos-contours}.

\subsection{The Super-Kamiokande Experiment} \label{sec:superk}

Super-Kamiokande is a multi-purpose experiment that has made significant contributions using both solar and atmospheric neutrinos as well as searching for proton decay. Running since 1996 it has collected the largest sample of atmospheric neutrinos to date. The experimental apparatus is a 50-kton water Cherenkov detector located in Japan at a depth of 2700 m.w.e. This detector consists of two optically separated detector regions. The inner detector volume is created inside the tank at 2.5~m from the wall by a supporting structure for the inward and outward facing PMTs.  The internal detector surface area originally had a photocathode coverage of 40\% given by 11,146 inward-facing PMTs of 20~inch diameter. The outward facing PMTs are optically isolated from the inner array and consists of 1,885 Hammamatsu of 8~inch  PMTs. Details of the detector can be found in Ref.~\cite{ref:super-k-nim}. The neutrino interactions occurring in the inner detector are determined to be fully contained if there is no activity in the outer detector.  A 22.5-kiloton fiducial volume in the inner detector is defined for neutrino interactions.

Super-Kamiokande has been running since 1996 and thus has gone through several phases of data-taking.  The first phase (SK-I) spanned 5 years. During refilling operations for a second phase of the experiment, a cascade of implosions destroyed half of the PMTs of the detector. After this event, the experiment was restarted with a reduction of the original coverage down to 19\%. Data taking during that phase was approximately 3 years (SK-II). In 2006, the original photo coverage was restored resulting in the third phase of the experiment (SK-III). In 2008, the front-end electronics, data acquisition system and software trigger have been updated which defines the fourth phase (SK-IV) of data taking. 

The total data exposure collected for all phases is 220 kton-year. This results in a very high statistics sample of atmospheric neutrinos which can be divided into a variety of data sets: fully contained (FC), partially contained (PC) and upward-going muons (UP $\mu$). Events that deposit all of their energy in the inner detector are FC, while those originating in this region and exiting are PC. Neutrino interactions occurring in the rock beneath the detector produce muons that transverse the detector (through-going) or stop in the detector (stopping) and are thus classified as UP $\mu$. The fully contained events are further divided into sub-Gev and multi-GeV samples based on their visible energy with a 1.33~GeV cut-off. Events are classified further into $\mu$-like and $e$-like samples by the ring patterns of the single or most energetic ring. More details about the analysis can be found in Ref.~\cite{ref:superk-atmo}.

\begin{figure}[t]
\centering
\includegraphics[width=82mm]{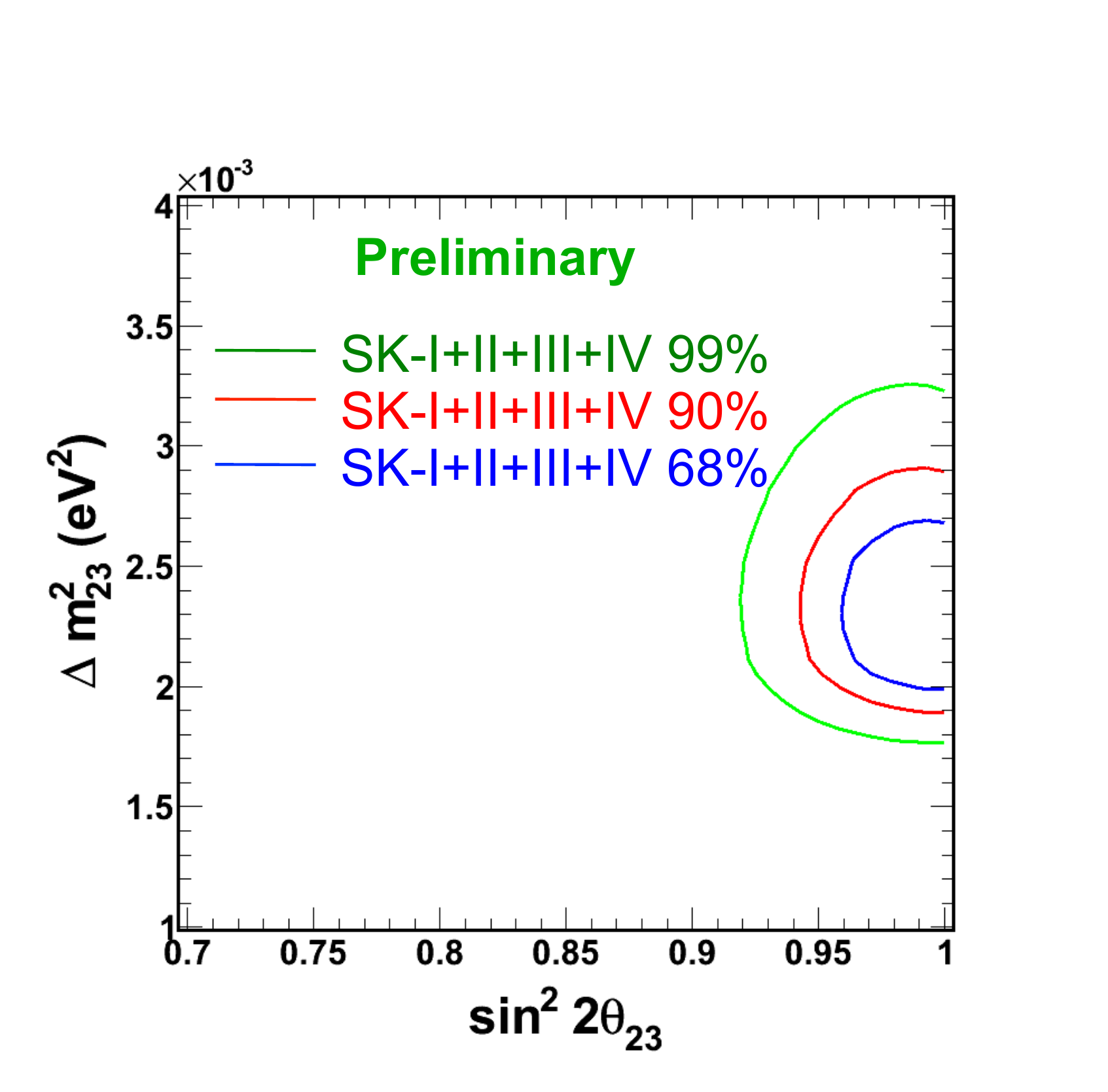}
\caption{Allowed neutrino oscillation parameters for the Super-Kamiokande (SK-I to SK-IV) data set. The 68\%, 90\%, and 99\% allowed region appear in thin (blue), medium (red), and thick  (green) lines, respectively.} \label{fig:sk-contour}
\end{figure}

Combining all data samples from SK-I to SK-IV, the Super-Kamiokande experiment achieves the World's best measurement in the mixing angle, resulting in $\rm sin^2\!(2\theta_{23})  > 0.96$  at 90\%. For the mass-squared difference, it gets $|\Delta m_{32}^2| = (2.2^{+0.15}_{-0.15})\times10^{-3}$\,eV$^2$.  The resulting 68\%, 90\% and 99\%\,C.L. contours are shown in Fig.~\ref{fig:sk-contour}. These results are in excellent agreement with the MINOS experiment.

\begin{figure*}[t]
\includegraphics[width=16pc,height=12pc,scale=0.8]{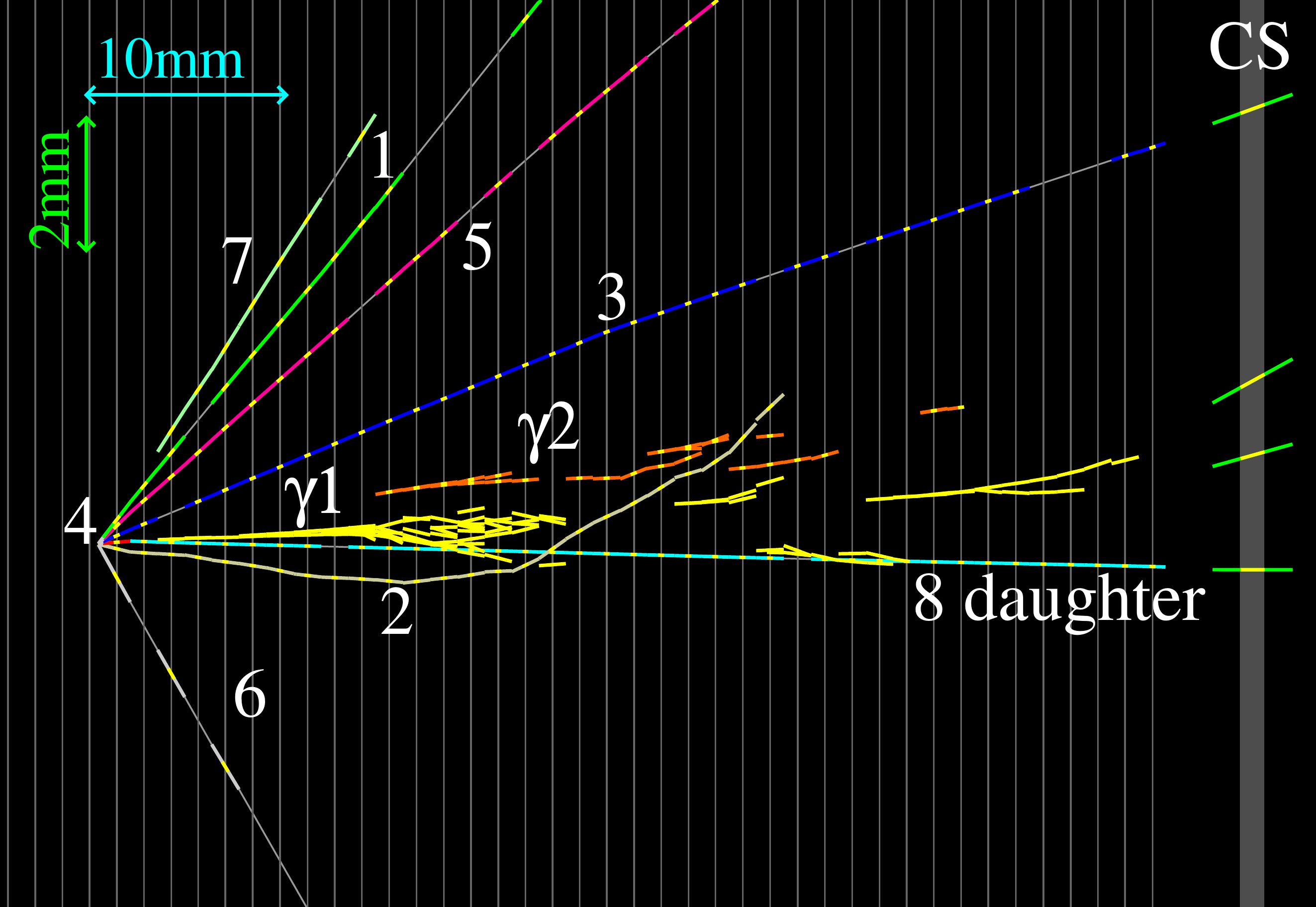}
\includegraphics[width=16pc,height=12pc,scale=0.8]{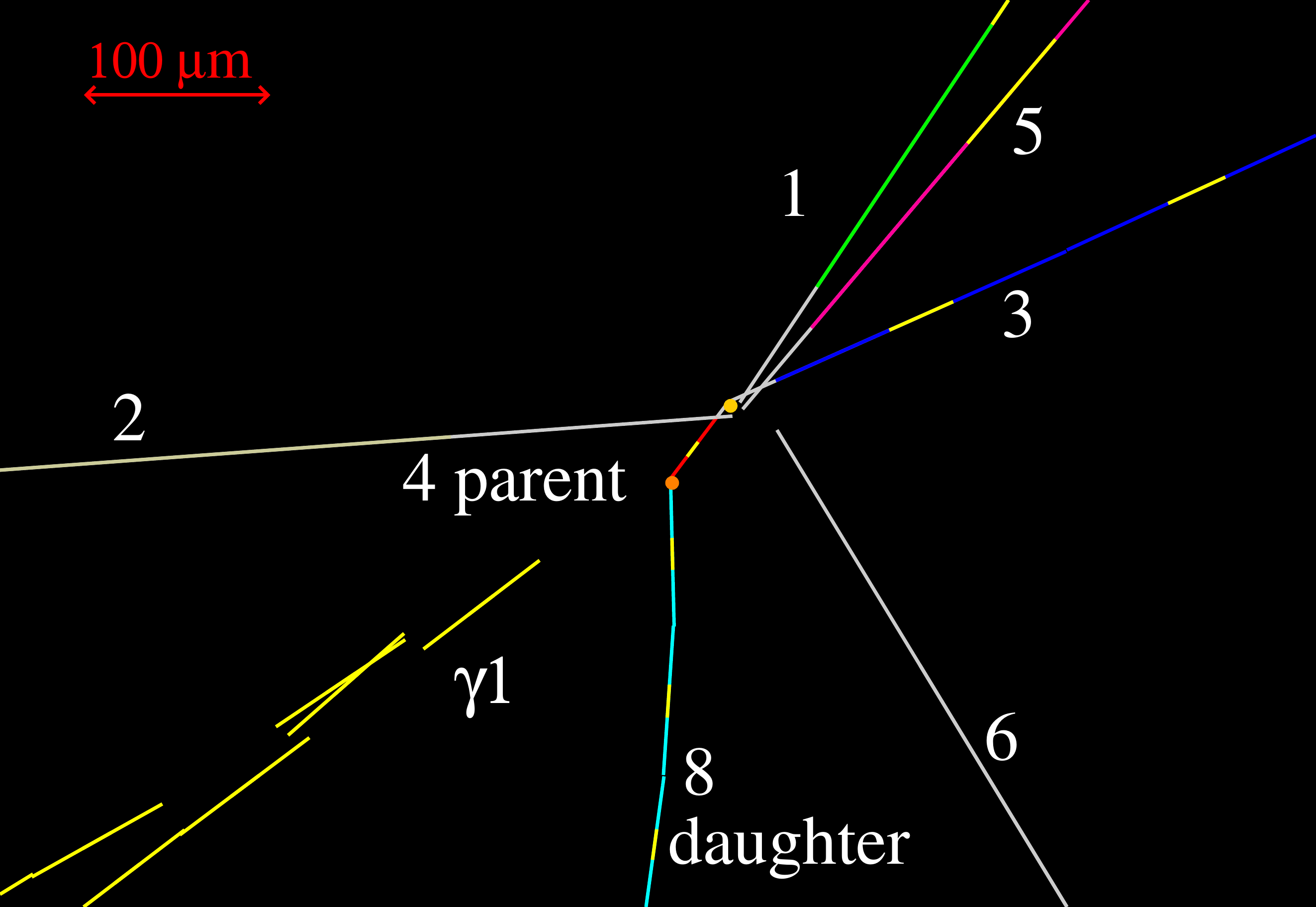}
\caption{Left: Longitudinal view
of the \nutau\ candidate event in OPERA. Right: Transverse view of the \nutau\ candidate event zoomed on the vertices.The short track named ``4 parent'' is the $\tau$
candidate.}
\label{fig:tau}
\end{figure*}

\section{TAU NEUTRINO APPEARANCE}

\subsection{The OPERA Experiment} \label{sec:OPERA}

The OPERA experiment distinguishes itself by aiming to perform the first direct observation of \numunutau\ oscillations by searching for \nutau{} appearance in a \numu\ beam. This experiment uses the CNGS beam at CERN which is created from 400~GeV protons and measured 730km away at the Grans Sasso Laboratory in central Italy. The neutrino beam energy ranges between 5 and 25~GeV so that it is higher than the 3.5~GeV kinematic threshold for \nutau\ production. This range is also significantly above the oscillation maximum at energy of 1.6~GeV, which results in a relatively low number of expected tau neutrino interactions. The beam contamination of \anumu\ is 2.1\%, and the \nue+\anue\ is lower than 1\%. In OPERA the neutrinos interact in a large mass target of lead plates that are alternated with emulsion films which provide the spatial resolution necessary to observed the short decay of the tau particle. More details of the detector design have been described elsewhere~\cite{ref:opera-nim}.

The observation of the first \nutau\ candidate event from \numunutau\ oscillations by the OPERA experiment was reported in 2010 for a data set corresponding to $1.89 \times 10^{19}$ POT. This event is shown in Figure~\ref{fig:tau} and it is compatible with being $\tau \rightarrow \rho^- \nutau$.  Details of the data selection procedures can be found in Ref.~\cite{ref:opera-tau}. The OPERA collaboration has now analyzed 92\% of the data sample from 2008-2009  with $4.8 \times 10^{19}$ and added significant improvements to the analysis and simulation chain, including better charm cross sections and reduction of the larger-than-expected charm background. No new \nutau\ candidate events are observed in the larger data set. In this data set, the observation of one event is compatible with the expectation of $1.65\pm 0.41$ total signal events and $0.16\pm0.03$ total background. The corresponding values specific to this channel ($\tau \rightarrow h$) are 0.49 for signal and $0.05\pm0.01$ for background. The probability of a background fluctuation is 15\% and the significance of observation is 95\%. OPERA has confirmed with a high degree of confidence that the dominant oscillation mode in this regime is to \numunutau. For the future, OPERA has already collected more than $7 \times 10^{19}$ POT in the 2010-2011 period which is currently  being analyzed.

\subsection{The Super-Kamiokande Experiment}

As described in Sect.~\ref{sec:superk}, the Super-Kamiokande experiment has accumulated a very large sample of atmospheric neutrinos since the start of running in 1996. However the detection of tau neutrino interaction in this detector remains challenging for a few reasons. First, the neutrino energy threshold for tau lepton production is 3.5~GeV, above which the atmospheric neutrino flux is relatively low. Second, the resulting event topology after the tau lepton decays is complicated ($i.e.$ multi-rings are likely) and the leading lepton might not be distinguishable. In addition, kinematic signatures cannot be used since the incident neutrino direction is not known a priori. Thus Super-Kamiokande employs likelihood and neural network techniques to discriminate tau neutrino events from atmospheric neutrino events on a statistical basis. Further details on this analysis can be found in Ref.~\cite{ref:superk-tau} and in a recent update to this analysis~\cite{Wendell:2011zz}. The new analysis includes data from the SK-I to SK-III phases of the experiment. 

\begin{figure}[t]
\centering
\includegraphics[width=70mm]{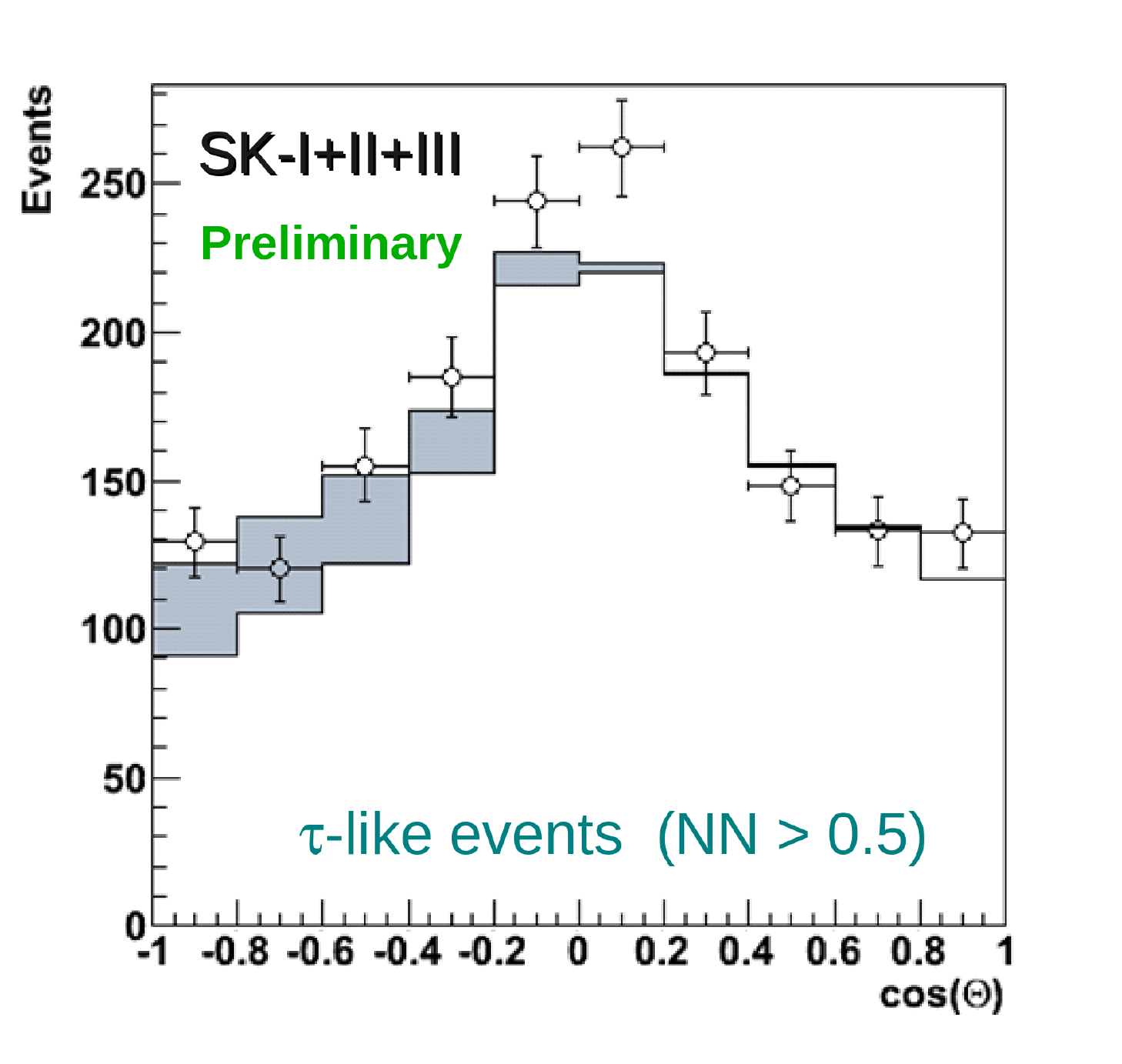}
\caption{Super-Kamiokande's zenith angle distribution for selected tau-like events in the SK-I to SK-III data set. The data sample is fitted after \nutau\ event selection criteria are applied. The histogram shows the best fit including \nutau\ (gray) and the background (white) from atmospheric neutrinos. A fitted excess of tau-like events in the upward-going direction is shown in the shaded area. } \label{fig:sk-tau}
\end{figure}

The analysis searches for an excess over background which must be oscillation induced.  A tau-enriched sample is selected using a neural net with 80\% efficiency. The zenith angle distribution for this sample is fitted with a combination of the expected tau neutrino signals resulting from oscillations and the predicted atmospheric neutrino background including oscillations. The data and best fitted distribution for signal and background are shown in Figure~\ref{fig:sk-tau}. An excess from the \nutau\ signal clearly appears in the upward-going region. The fitted zenith angle distribution is parametrized as follows: $Data = \alpha \times bkg + \beta \times signal$ where the sample normalizations $\alpha$ and $\beta$ are allowed to vary freely. The normalization of the best fit for the background in the likelihood fit is $\beta = 1.63 \pm 0.35$ corresponding to 213.6 \nutau\ events. The data disfavors the absence of tau neutrino appearance at 3.8$\sigma$ thus adding evidence to the claim that the dominant mode in this regime is indeed \numunutau\ oscillations.

\section{MUON ANTI-NEUTRINO DISAPPEARANCE}

\subsection{The MINOS Experiment}
By inverting the current of the horns and reversing the polarity of the detector magnets, the conditions of the MINOS experiment are optimized for the anti-neutrino disappearance measurement. This enables MINOS to perform a comparison of the values of parameters governing oscillations of neutrinos with the values for antineutrino oscillations, expected to be identical in the standard picture. 

In Sect.~\ref{sec:MINOS} we mentioned that the NuMI beam is capable of running in an anti-neutrino dominant mode. In this mode, the \anumu-enhanced beam has a peak energy of 3~GeV. The fraction of \numu-interactions is about 21\% below 6~GeV and rises up to 81\% at 20~GeV (for no oscillations).  MINOS has collected a data set of $2.95 \times 10^{20}$~POT running in this mode. Using similar methods for the data selection to those used in the MINOS \numu\ disappearance analysis, a sample of \numu\ and \anumu\,CC-like events is obtained. By exploiting the magnetized detectors, a sample with positively charged tracks can be easily identified. In the oscillation region, the resulting \anumu\ sample has purity of 98\%, with the background equally divided between neutral current events and \numu-CC events. In the same region the selection has an efficiency of 96\%. Further details of the analysis can be found in Ref.~\cite{ref:minos-antinu}. 

\begin{figure}[t]
\centering
\includegraphics[width=80mm]{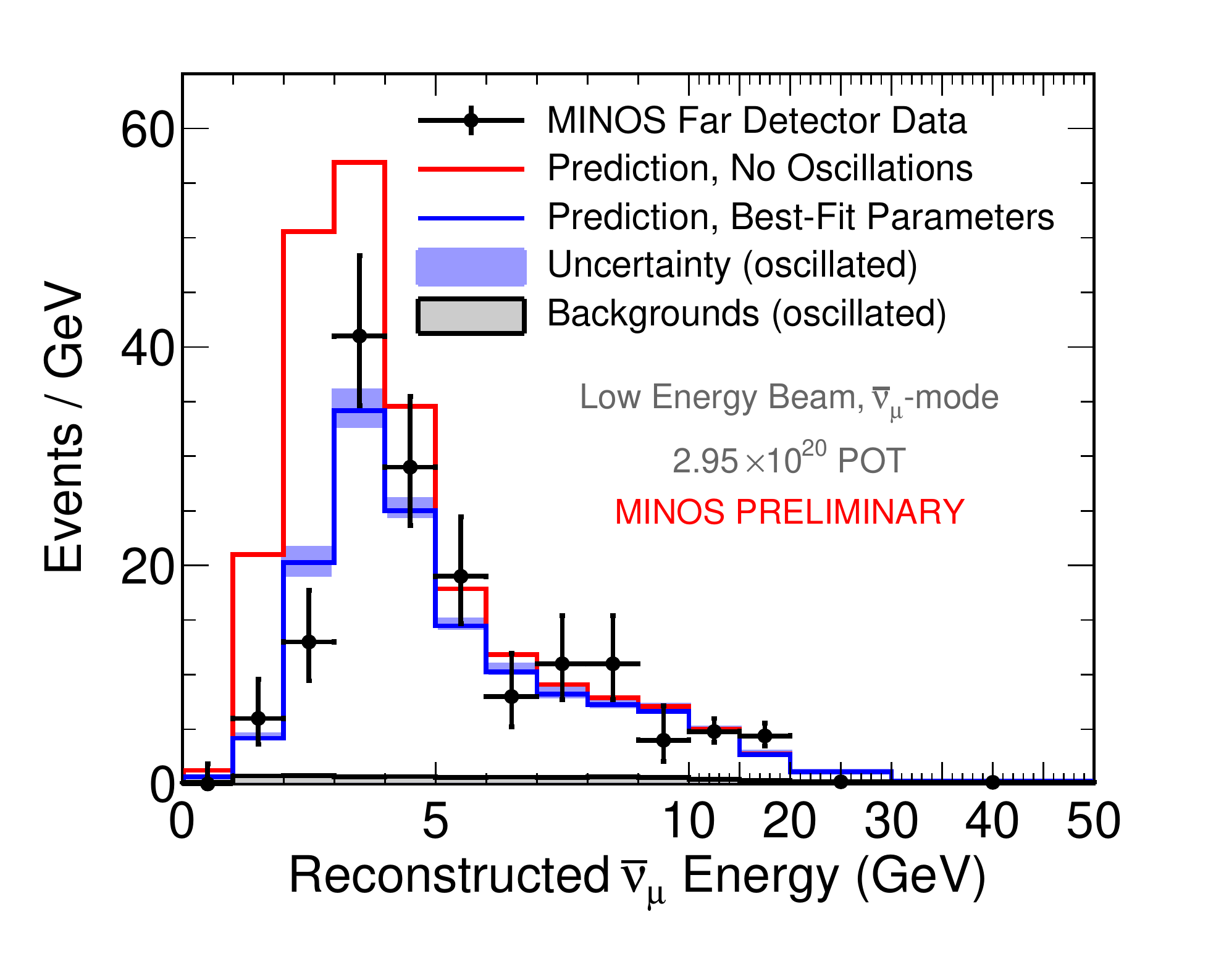}
\caption{Comparison of the measured  \anumu\ CC energy
spectrum  from MINOS to the expectation in the absence of
oscillation (red) and using the oscillation parameters which best fit to the 
\anumu\ data (blue). The total expected background is also
indicated.} \label{fig:minos-anu-sp}
\end{figure}

Using the prediction obtained from the near detector data, we expect 273 selected \anumu\ CC events with energy below 50~GeV  in the MINOS far detector in the absence of oscillation and we observe 193 events in the data.  Figure~\ref{fig:minos-anu-sp} shows the energy spectra of these events presenting a clear energy dependent deficit. This is unique evidence for \anumu\ disappearance consistent with oscillation in a \anumu-tagged sample. The no-oscillation hypothesis is disfavored at 7.3$\sigma$ by MINOS.

\begin{figure}[t]
\centering
\includegraphics[width=80mm]{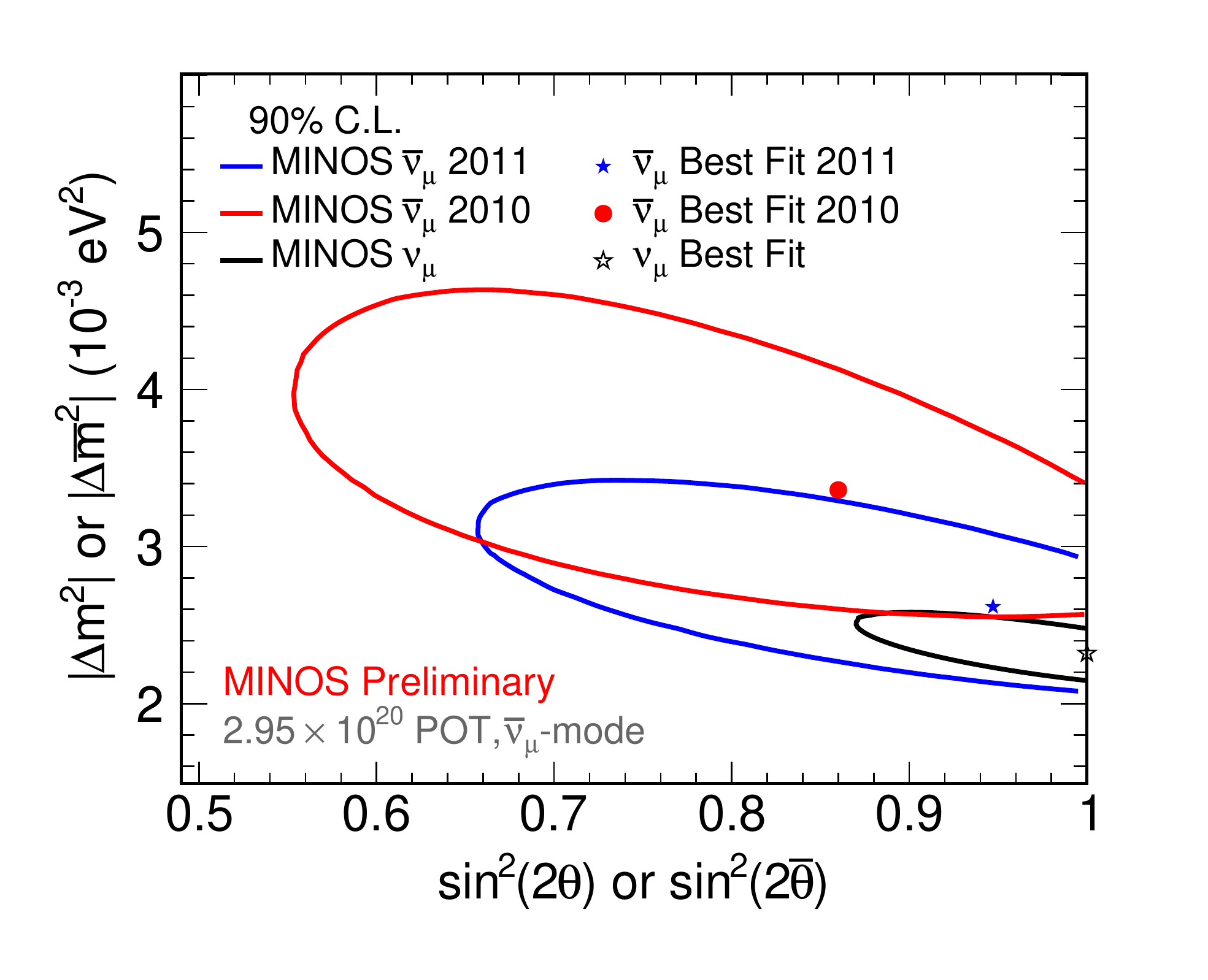}
\caption{MINOS allowed regions for the \anumu\ (blue) and \numu\ (black) oscillation parameters from
a fit to the data, including all sources of systematic uncertainty. The MINOS allowed region for  \anumu\  oscillation published prior to this analysis~\cite{ref:minos-antinu} is also shown (red).} \label{fig:minos-anu-contour}
\end{figure}

Comparing the prediction to the data energy spectrum using a binned log likelihood, the following oscillation parameters are found:  $|\Delta \bar{m}_{32}^2| = (2.62^{+0.31}_{-0.28})\times10^{-3}$\,eV$^2$ and $\rm sin^2\!(2\bar{\theta}_{23})  > 0.75$. The 90\% CL limit on the oscillation parameters are shown in Figure~\ref{fig:minos-anu-contour}. As can be seen in both Figures~\ref{fig:minos-anu-sp} and \ref{fig:minos-anu-contour}, these results are compatible with those obtained for neutrino oscillations ($p=42$\%). MINOS has since collected more anti-neutrino data which is being analyzed at the time of this writing. 

\subsection{The Super-Kamiokande Experiment}

Atmospheric neutrino data carries both neutrinos and anti-neutrinos; however Super-Kamiokande, a water Cherenkov experiment, cannot distinguish on an event-by-event basis between them. Fortunately, potential differences in neutrino and anti-neutrino oscillations can appear in a large atmospheric neutrino sample statistically. The key is that the relative numbers of the neutrinos versus anti-neutrinos are expected to be  different for several reasons: their cross-sections differ by a factor of 2 to 3 depending on neutrino energy, the ratio of their fluxes is also energy dependent and kinematic consideration induce difference in the products of their reactions. Thus oscillations for neutrinos and anti-neutrinos could differ and this would modify the zenith angle distribution to account for different characteristic energies and path lengths of the oscillation of each species. Since Super-Kamiokande does observe maximal oscillations for the sum of neutrinos and anti-neutrinos, this places a strong constraint on the allowed oscillation parameters of anti-neutrinos. 

The analysis follows similar data breakdown and selection procedure as described in Sect.~\ref{sec:superk} and a more detailed description can be found in  Ref.~\cite{ref:superk-antinu}.
Anti-neutrino oscillations are considered independently of neutrino oscillations over a four-dimensional oscillation space with two parameters for each: 
($\Delta \bar m_{32}^{2}$, $\mbox{sin}^{2} 2\bar{\theta}_{23}$) and 
($\Delta m_{32}^{2}$, $\mbox{sin}^{2} 2\theta_{23}$). Additionally, there are 120 sources of systematic uncertainty considered. The best fit point 
($\Delta \bar m_{32}^{2}$, $\mbox{sin}^{2} 2\bar{\theta}_{23}$) = ( $ 2.0 \times 10^{-3} \mbox{eV}^{2}$, $1.0$~) is consistent with that of that of the neutrino analysis. While Super-Kamiokande does not pose a more stringent constraint than MINOS on \delmsq{32}, the mixing angle is further constrained to $\rm sin^2\!(2\bar{\theta}_{23})  > 0.83$ at 90\% CL.

\begin{figure}[t]
\centering
\includegraphics[width=80mm]{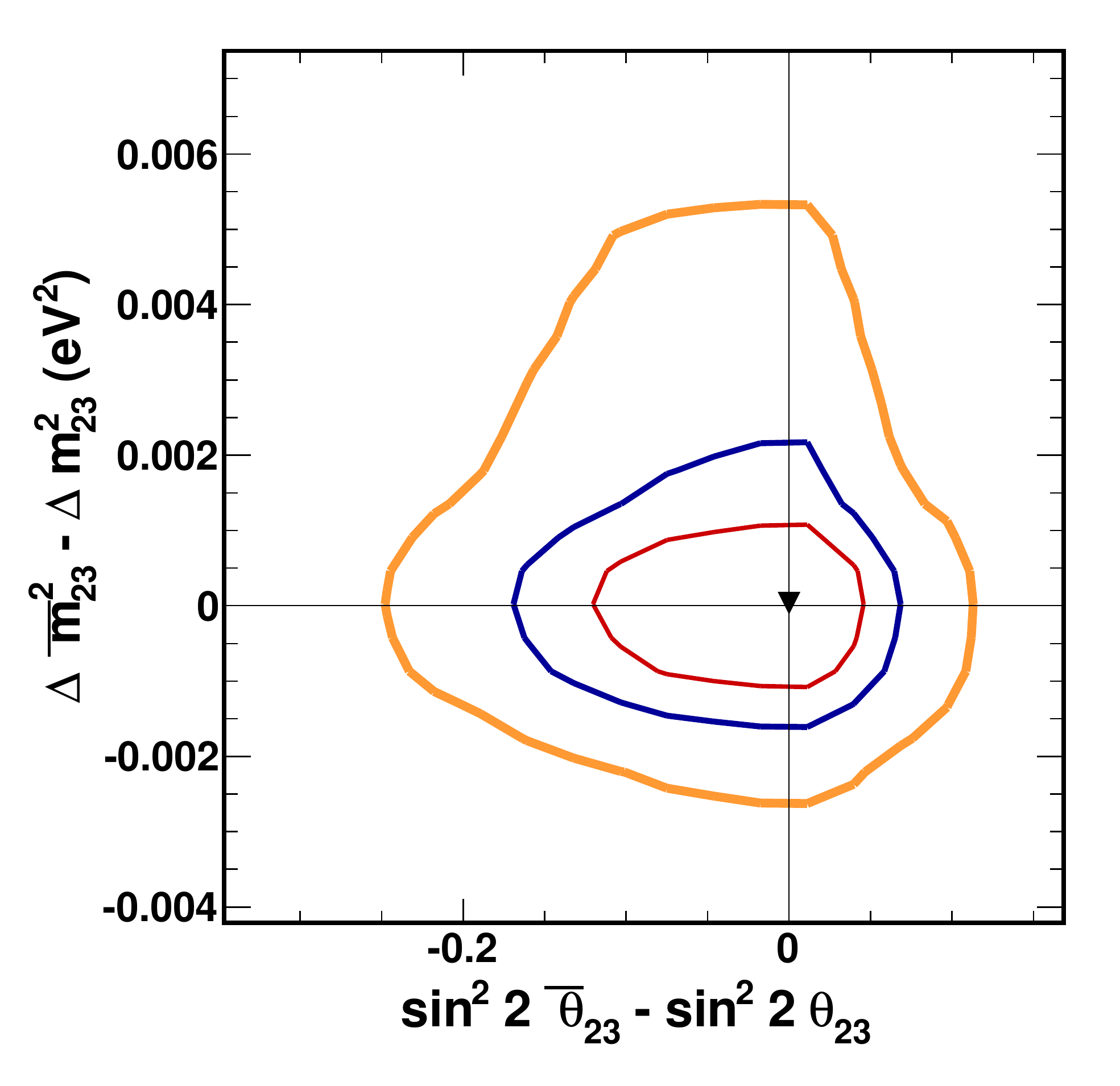}
\caption{Super-Kamikande's allowed differences between neutrino and antineutrino oscillation parameters using the SK-I to SK-III data set. The 68\%, 90\%, and 99\% allowed region appear in thin (red), medium (blue), and thick (orange) lines, respectively.} \label{fig:sk-anu}
\end{figure}

Figure~\ref{fig:sk-anu} illustrates the difference between neutrino and antineutrino oscillations 
permitted by the data. This figure shows the allowed  regions as a function of the difference of the antineutrino and  neutrino mixing angles, $\mbox{sin}^{2}2\bar \theta_{23} - \mbox{sin}^{2}2 \theta_{23}$, and mass squared splittings, $\Delta \bar m_{32}^{2} - \Delta m_{32}^{2}$. 
The black triangle near the origin represents the position of the best fit. 

\section{OFF-AXIS MUON NEUTRINO DISAPPEARANCE}

\subsection{The T2K Experiment}

The T2K experiment is designed with the primary goal of searching for \nue\ appearance over a 295~km baseline. It uses a neutrino beam from the JPARC facility in Japan that starts with 30 GeV protons and has a design intensity of 750 kW. The beam is aimed at Super-Kamiokande, the large water Cherenkov detector that first saw evidence for atmospheric neutrino oscillation and was previously described. This detector is 2.5$^\circ$ off the beam axis which results in a narrow beam profile peaked at the oscillation maximum. A near detector complex with detectors both on and off axis provides measurement of the beam profile and direction as well as the beam flux, cross sections and beam composition.  More information about the experimental design can be found in these proceedings~\cite{ref:karlen} and elsewhere~\cite{ref:t2k-nim}

The T2K beam data taking started in January 2010 and was interrupted in March 2011 due to an earthquake in Eastern Japan. The analysis carried out for the muon neutrino disappearance is based on $1.43 \times 10^{20}$~POT. During this time the beam power reached 145 kW. 

The data selection for this analysis is based on that of the Super-Kamiokande detector but restricted to select only fully contained charged-current quasi-elastic processes in the water which are identified by a clean single ring from the leading muon. The main background are charged current processes with one charged pion in the final state. Other selection criteria include requiring events to have less than 2 decay electrons and a reconstructed muon larger than 200~MeV$/c$. After this selection criteria 31 pass all selections (from 33 that have a single mu-like ring). Under the null oscillation hypothesis the expected number is  $104 ^{+14}_{-13}$ rejecting the null hypothesis at 4.5$\sigma$.

\begin{figure}[tb]
\includegraphics[width=80mm]{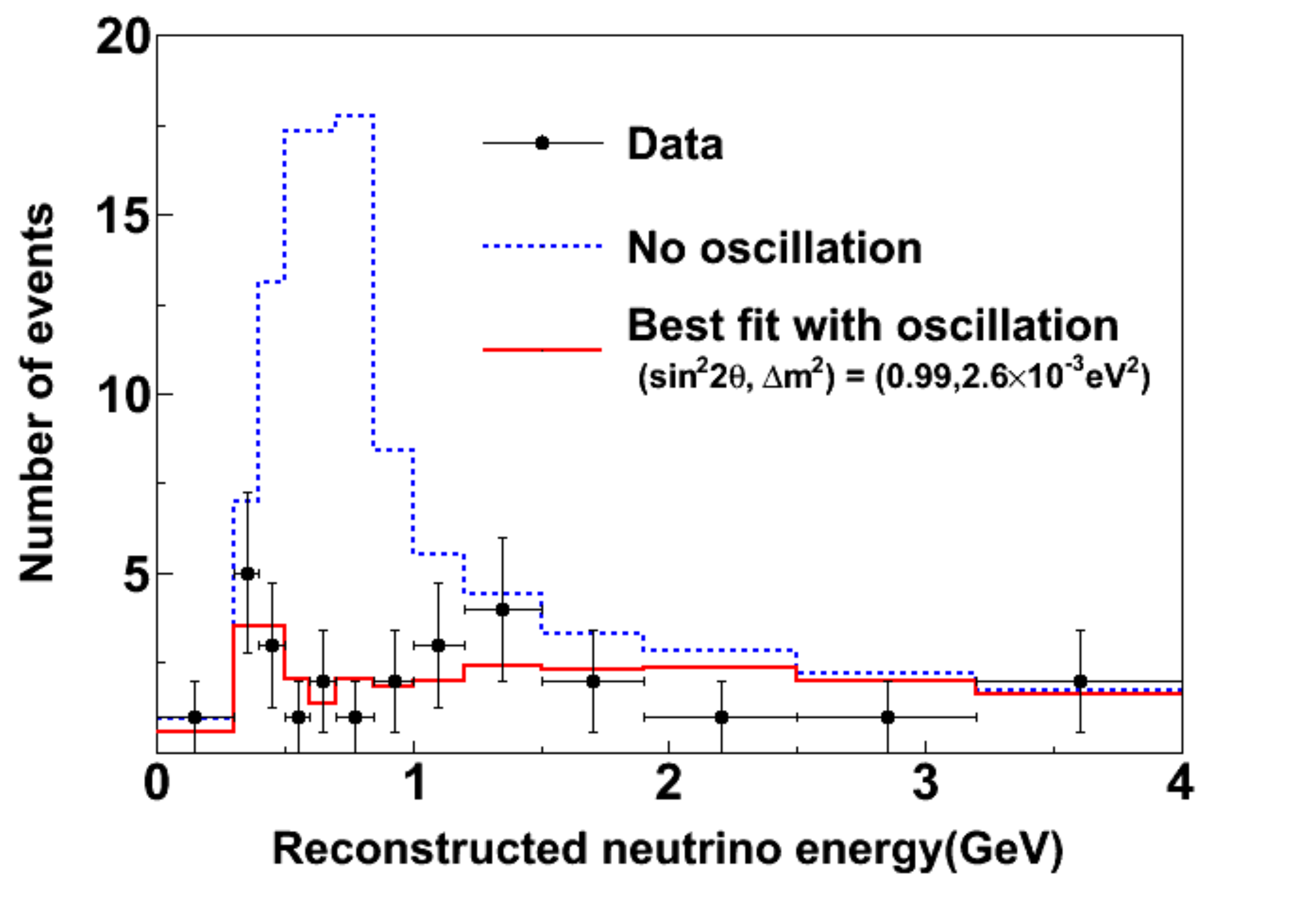}
\includegraphics[width=80mm]{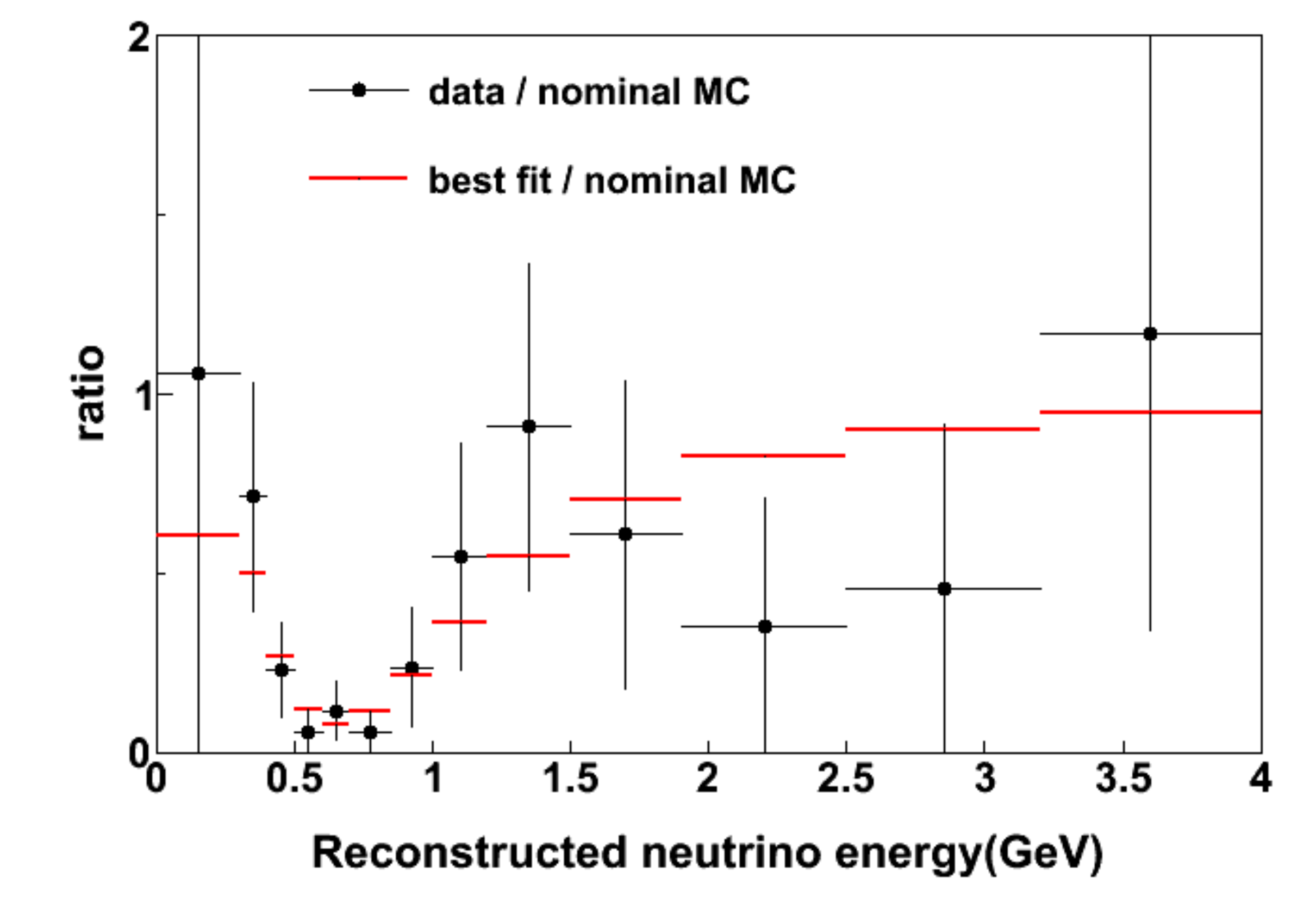}
\caption{T2K's reconstructed neutrino energy spectrum at the far detector (top) and data/MC ratio (bottom) for muon neutrino events~\cite{ref:t2k-proc}} 
\label{fig:t2k-numu}
\end{figure}

The reconstructed neutrino energy spectrum demonstrates the strength of the off-axis concept. With the beam peak centered around the oscillation maximum, just a few events can show a clear oscillation pattern as it is highlighted by the data/MC ratio in Figure~\ref{fig:t2k-numu}.

\subsection{The NOvA Experiment}

The NOvA experiment currently under construction is also designed to search for \numunue\ oscillations by comparing electron neutrino rates at Fermilab to those observed in northern Minnesota at 810~km from the target with a large 14-kton detector located on the surface.  By placing the detector 14~mrad off-axis of the NuMI beam line, the neutrino flux is enhanced as a narrow peak at 2~GeV (the first oscillation maximum),  offering a significant reduction of background from high energy neutral current neutrino interactions. The loss in flux due to being off-axis is compensated by upgrades of the NuMI beam to 700~kW. 

The NOvA detectors are totally active tracking calorimeters, optimized for the identification of electron neutrino interactions, with fine sampling of the characteristic electromagnetic showers. Complete details about the detector design are described in the experiment's Technical Design Report~\cite{ref:nova-tdr}.

\begin{figure}[t]
\centering
\includegraphics[width=80mm]{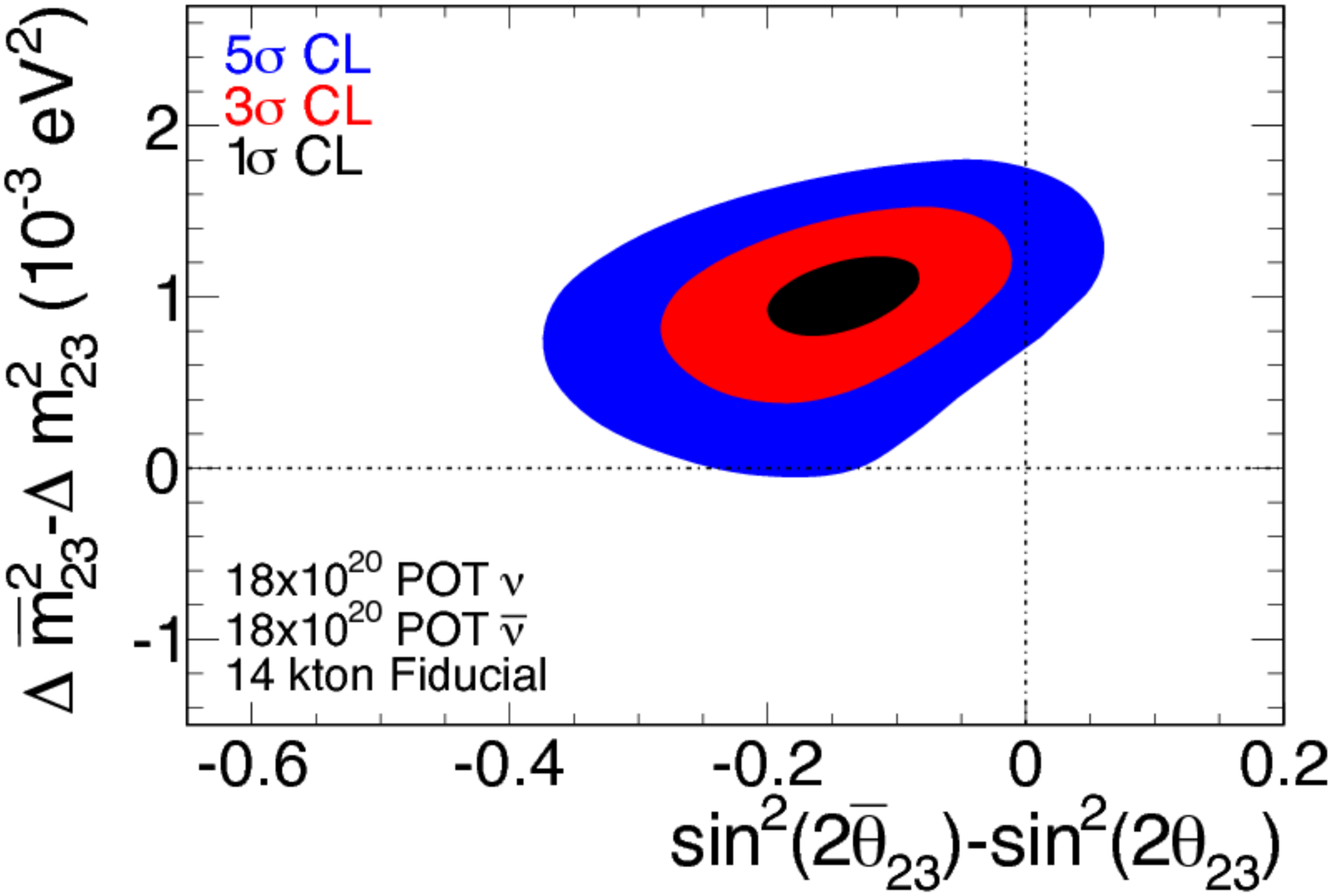}
\caption{NOvA projected sensitivity for the difference in neutrino and anti-neutrino muon neutrino disappearance parameters. The difference is set to coincide with the previously reported value by MINOS \cite{ref:minos-antinu}. Contours are based on three years of neutrino running and three years of anti-neutrino running and use quasi-elastic events.} \label{fig:nova}
\end{figure}

NO$\nu$A will also significantly improve the precision on the measurements of atmospheric neutrino oscillation parameters $\Delta m^2_{32}$ and $\theta_{23}$ down to 1-2\%. For an asymmetry between $\Delta m^2_{32}$ and $\Delta \bar{m}^2_{32}$ such as the one hinted previously by MINOS~\cite{ref:minos-antinu}, NO$\nu$A could establish this difference at the 5$\sigma$ level after the full 6 year run, 3 + 3 years in neutrinos and antineutrinos as shown in Figure~\ref{fig:nova}. While MINOS neutrinos and anti-neutrinos are now compatible at the 42\% level, this still illustrates the capabilities of NOvA in this respect. 

\section{CONCLUSIONS}

\begin{figure}[t]
\centering
\includegraphics[width=80mm]{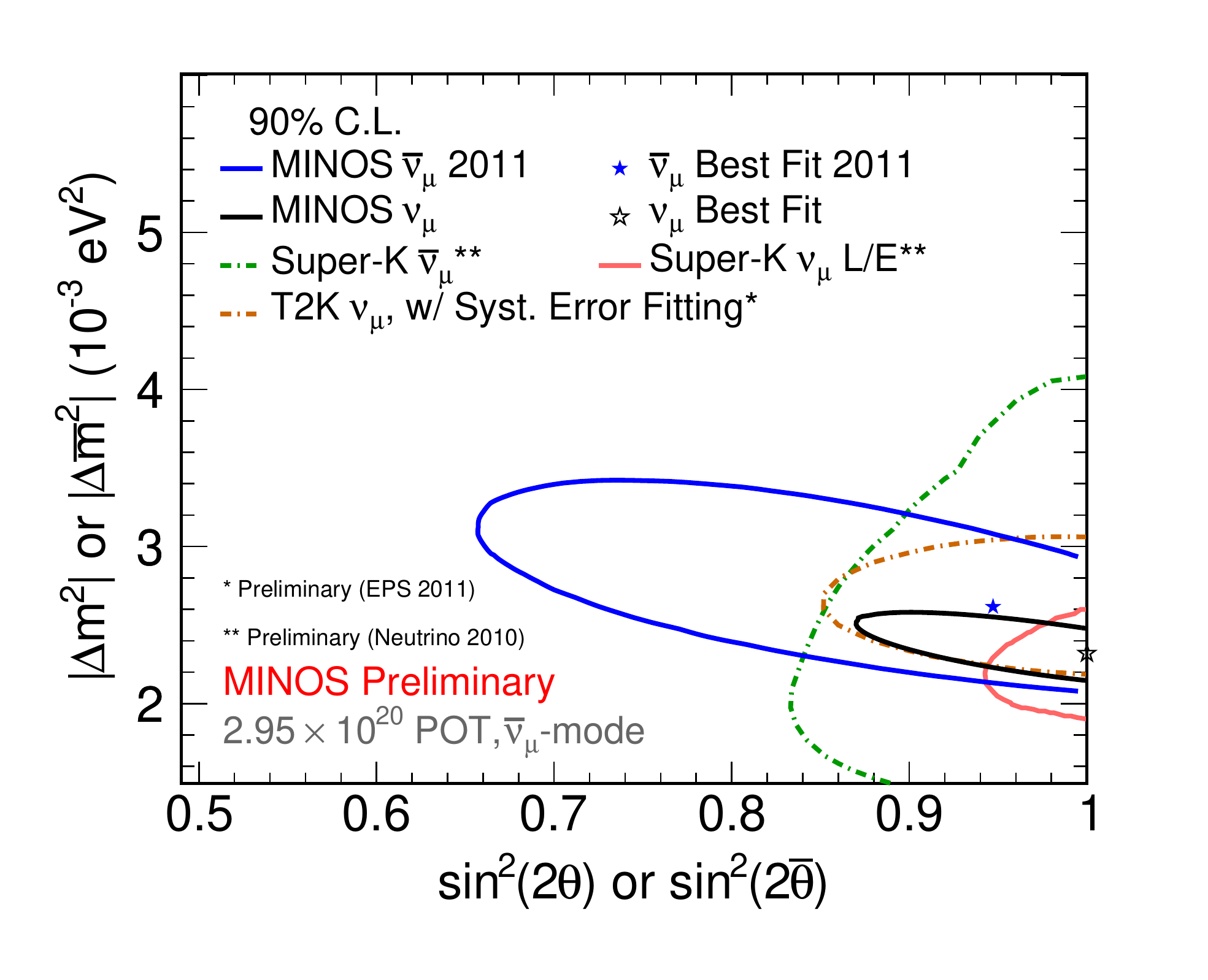}
\caption{Allowed regions at 90\% CL for the \numu\ and  \anumu\ oscillation parameters from each experiment in these proceedings.} \label{fig:combined-contour}
\end{figure}

Over the last decade, long-baseline and atmospheric neutrino experiments have significantly narrowed the uncertainties and degrees of freedom in the muon-tau sector of neutrinos.  The dominant mode of oscillation from muon neutrino is demonstrated to be tau neutrinos: one event has been observed directly in Opera and many more indirectly by Super-Kamiokande. Similarly, the parameter space for anti-neutrinos is being explored aggressively by MINOS and Super-Kamiokande.  Figure~\ref{fig:combined-contour} shows our best knowledge of the oscillation parameters for this sector both for neutrinos and anti-neutrinos as of this writing. While it is clear that we have converged in a very consistent picture of this sector, there can still be plenty of surprises hidden around the corner. The up and coming generation of off-axis long-baseline experiments are well under way to make very precise measurements in this sector. We eagerly await future results from T2K's additional data, NOvA first data and more data from MINOS and Super-Kamiokande.

%

\bigskip 
\bibliography{basename of .bib file}

\end{document}

%% file: mycmds.tex
\newcommand{\delmsq}[1]{\ensuremath{\Delta m^2_{ #1 }}}
\newcommand{\dmsq}[1]{\delmsq{ #1 }}

\newcommand{\tonetwo}{\ensuremath{\theta_{12}}}
\newcommand{\tonethree}{\ensuremath{\theta_{13}}}
\newcommand{\ttwothree}{\ensuremath{\theta_{23}}}

\def\numunue{$\nu_\mu \rightarrow \nu_e$}
\def\numunumu{$\nu_\mu \rightarrow \nu_{\mu}$}
\def\numunutau{$\nu_\mu \rightarrow \nu_{\tau}$}

\newcommand{\numu}{\mbox{$\nu_{\mu}$}}                   
\newcommand{\nue}{\mbox{$\nu_{e}$}}                      
\newcommand{\nutau}{\mbox{$\nu_{\tau}$}}                 
\newcommand{\anue}{\mbox{$\overline{\nu}_{e}$}}          
\newcommand{\anumu}{\mbox{$\overline{\nu}_{\mu}$}}       

\newcommand{\ket}[1] {| #1 \rangle}